\documentclass[12pt]{article}

\title{
JACKSON'S $q$-EXPONENTIAL AS THE EXPONENTIAL OF A SERIES}
\author{C. QUESNE\\
{\small \sl Physique Nucl\'eaire Th\'eorique et Physique Math\'ematique,}\\ {\small \sl
Universit\'e Libre de Bruxelles, Campus de la Plaine CP229,} \\ {\small \sl  Boulevard~du
Triomphe, B-1050 Brussels, Belgium} \\
{\small \sl E-mail: cquesne@ulb.ac.be}}
\date{ }
\begin{document}
\baselineskip=22pt plus 1pt minus 1pt
\maketitle

\begin{abstract} 
Jackson's $q$-exponential is expressed as the exponential of a series  whose coefficients
are obtained in closed form. Such a relation is used to derive some properties of the
$q$-exponential.
\end{abstract}

\noindent
Running head: Jackson's $q$-Exponential

\noindent
Keywords: Basic special functions; $q$-exponential

\noindent
PACS Nos.: 02.20.Uw, 02.30.Gp
\par
%
%
\bigskip
Since the advent of quantum groups, Jackson's $q$-exponential~\cite{jackson} has
found a lot of applications in physics. As an example, we may quote its use to describe
some generalized coherent states in quantum optics~\cite{arik, cq}.\par
%
%
Properties of the $q$-exponential are often derived from its expression as an infinite
product~\cite{ubriaco}. For practical purposes, however, it may be useful to know the
link between the $q$-exponential and an elementary function, such as the ordinary
exponential. It is the purpose of the present letter to establish such a relation.\par
%
%
Jackson's $q$-exponential, defined by \footnote{Jackson has actually introduced two
different kinds of $q$-exponentials, but the other one is related to the inverse of that
considered here.}
\begin{equation}
  E_q(z) = \sum_{k=0}^{\infty} \frac{z^k}{[k]_q!}, 
\end{equation}
where
\begin{equation}
  [k]_q!  \equiv  \left\{\begin{array}{ll}
        1 & {\rm if\ } k=0, \\[0.2cm]
        [k]_q [k-1]_q \ldots [1]_q & {\rm if\ } k=1, 2, \ldots,
     \end{array}\right.  
\end{equation}
and 
\begin{equation}
  [k]_q \equiv \frac{1-q^k}{1-q} = 1 + q + q^2 + \cdots + q^{k-1},
  \label{eq:number}
\end{equation}
has a finite radius of convergence $[\infty]_q = (1-q)^{-1}$ if $0 < q < 1$, but
converges for all finite $z$ if $q > 1$~\cite{exton}. In the appropriate region of definition,
let us try to express it as the exponential of some series,
\begin{equation}
  E_q(z) = \exp\left(\sum_{k=1}^{\infty} c_k(q) z^k\right). \label{eq:result}
\end{equation}
This amounts to expanding in Taylor series the logarithm of the $q$-exponential,
\begin{equation}
  \ln E_q(z) = \sum_{k=1}^{\infty} c_k(q) z^k. 
\end{equation}
\par
%
%
In Ref.~\cite{pourahmadi}, it has been shown that if the functions $f(z) =
\sum_{k=0}^{\infty} a_k z^k$ and $h(z) = \ln f(z) = \sum_{k=1}^{\infty} c_k z^k$,
where $a_0=1$, are analytic in some neighbourhood of zero, then the Taylor coefficients
of $h(z)$ satisfy the recursion relation
\begin{equation}
  c_k = a_k - \frac{1}{k} \sum_{j=1}^{k-1} j a_{k-j} c_j, \qquad k=2, 3, \ldots,
\end{equation}
with $c_1 = a_1$. Applying this result to the logarithm of the $q$-exponential leads to
the relations
\begin{eqnarray}
  c_k(q) & = & \frac{1}{[k]_q!} - \frac{1}{k} \sum_{j=1}^{k-1} \frac{j}{[k-j]_q!} c_j(q),
        \qquad k=2, 3, \ldots, \label{eq:recursion}\\
  c_1(q) & = & 1. \label{eq:condition}
\end{eqnarray}
\par
%
%
It is straightforward to check that the solution of Eq.~(\ref{eq:recursion}), satisfying
condition (\ref{eq:condition}), is provided by
\begin{equation}
  c_k(q) = \frac{(1-q)^{k-1}}{k [k]_q}, \qquad k=1, 2, 3, \ldots. \label{eq:solution}
\end{equation}
Inserting such an expression in Eq.~(\ref{eq:recursion}) converts the latter into the
relation
\begin{equation}
  \sum_{j=1}^k  
     \left[\begin{array}{c}
         k \\ j
     \end{array}\right]_q
  (1-q)^{j-1} [j-1]_q! = k, \qquad k=2, 3, \ldots, \label{eq:relation}
\end{equation}
where
\begin{equation}
  \left[\begin{array}{c}        
      k \\ j
  \end{array}\right]_q \equiv \frac{[k]_q!}{[j]_q!\, [k-j]_q!}
\end{equation}
denotes a $q$-binomial coefficient~\cite{exton}. Equation (\ref{eq:relation}) can be
easily proved by induction over $k$ by using the recursion relation
\begin{equation}
  \left[\begin{array}{c}        
      k \\ j
  \end{array}\right]_q 
  = q^j   \left[\begin{array}{c}        
                 k-1 \\ j
             \end{array}\right]_q
  + \left[\begin{array}{c}        
         k-1 \\ j-1
     \end{array}\right]_q, \qquad j=1, 2, \ldots, k-1.
\end{equation}
We indeed obtain
\begin{eqnarray}
  \lefteqn{\sum_{j=1}^k  
       \left[\begin{array}{c}                              
           k \\ j
       \end{array}\right]_q (1-q)^{j-1} [j-1]_q!} \nonumber\\
  & = & \sum_{j=1}^{k-1}  
       \left[\begin{array}{c}                              
           k-1 \\ j
       \end{array}\right]_q q^j (1-q)^{j-1} [j-1]_q!
       + \sum_{j=0}^{k-2}  
       \left[\begin{array}{c}                              
           k-1 \\ j
       \end{array}\right]_q (1-q)^j [j]_q! \nonumber \\
  && \mbox{} + (1-q)^{k-1} [k-1]_q! \nonumber \\
  & = & \sum_{j=1}^{k-1}  
       \left[\begin{array}{c}                              
           k-1 \\ j
       \end{array}\right]_q q^j (1-q)^{j-1} [j-1]_q!
       + 1 \nonumber \\
  && \mbox{} + \sum_{j=1}^{k-1}  
       \left[\begin{array}{c}                              
           k-1 \\ j
       \end{array}\right]_q (1-q)^{j-1} (1-q^j) [j-1]_q! \nonumber \\
  & = & \sum_{j=1}^{k-1}  
       \left[\begin{array}{c}                              
           k-1 \\ j
       \end{array}\right]_q (1-q)^{j-1} [j-1]_q! + 1 \nonumber \\
  & = & (k-1) + 1,
\end{eqnarray}
where in the last step use has been made of the induction hypothesis.\par
%
%
Equations (\ref{eq:result}) and (\ref{eq:solution}) are the central result of this paper. To
illustrate their usefulness, one can apply them to derive the following properties of the
$q$-exponential, already quoted in Ref.~\cite{ubriaco} (where we have corrected some
misprints),
\begin{equation}
  E_q(z) E_{q^{-1}}(z) = 1,
\end{equation}
\begin{equation}
  E_q(z) E_q(-z) = E_{q^2}\left(\frac{1-q}{1+q}z^2\right),  \label{eq:property}
\end{equation}
\begin{equation}
  E_q([n]_q z) = \prod_{m=0}^{n-1} E_{q^n}(q^m z), \qquad n=2, 3, \ldots,
\end{equation}
as well as a generalization of Eq.~(\ref{eq:property}),
\begin{equation}
  \prod_{m=0}^{n-1} E_q(e^{2\pi {\rm i}m/n} z) = E_{q^n}\left(\frac{(1-q)^{n-1}}
  {[n]_q} z^n\right), \qquad n=2, 3, \ldots.
\end{equation}
The proof of these relations is based upon the multiplicative property of the ordinary
exponential, $\exp(x) \exp(y) = \exp(x+y)$, and on some elementary properties of the
coefficients $c_k(q)$, defined in (\ref{eq:solution}),
\begin{equation}
  c_k(q^{-1}) = (-1)^{k-1} c_k(q),
\end{equation}
\begin{equation}
  2 c_{2k}(q) = \left(\frac{1-q}{1+q}\right)^k c_k(q^2),
\end{equation}
\begin{equation}
  [n]_{q^k} c_k(q^n) = ([n]_q)^k c_k(q),
\end{equation}
\begin{equation}
  n c_{nk}(q) = \left(\frac{(1-q)^{n-1}}{[n]_q}\right)^k c_k(q^n).
\end{equation}
\par
%
%
We think that the expression of Jackson's $q$-exponential as the exponential of a series
whose coefficients are known in closed form may find some interesting physical
applications. As a final point, it is worth noting that such a simple result does not seem to
hold true for the $q$-exponential defined in terms of symmetric $q$-numbers.\par
%
%
The author is indebted to K.\ A.\ Penson for informing her about Ref.~\cite{pourahmadi}
and for a valuable discussion. Thanks are also due to R.\ Jagannathan and K.\ Srinivasa
Rao for some interesting comments. The author is a Research Director of the National
Fund for Scientific Research (FNRS), Belgium.\par
%
%

\end{document}